\documentstyle[aps,pra,epsf,multicol]{revtex}

\newcommand{\eq}[1]{\begin{equation} #1 \end{equation}}
\newcommand{\eqa}[1]{\begin{eqnarray} #1 \end{eqnarray}}

\def\Tr{{\rm Tr}}
\def\Id{\textbf{1}}

\def\bin{\{0,1\}}
\def\bases{\{+,\times\}}

\newcommand{\ket}[2]{ | \, {#1} \rangle_{#2}}
\newcommand{\bra}[2]{\,_{#2} \langle {#1} \,  |}
\newcommand{\proj}[2]{\ket{#1}{#2}\bra{#1}{#2}}

\newcommand{\ketb}[1]{ | \, \underline{#1} \rangle}

\newcommand{\ketc}[1]{ | \, {#1} \rangle}

\newcommand{\scalarc}[2]{\langle {#1} \,  | \, {#2}  \rangle}

\def\state{|\psi\rangle}

\def\share{{\mbox{\scriptsize share}}}
\def\nshare{{\mbox{\scriptsize not share}}}

\def\E{{\mathcal{E}}}

\begin{document}

\title{Security of EPR-based Quantum Cryptography against Incoherent Symmetric Attacks}
\author{Hitoshi Inamori, Luke Rallan and Vlatko Vedral}
\address{Centre for Quantum Computation, Clarendon Laboratory, Oxford University}
\date {\today}
\maketitle
\begin{abstract}
We investigate a new strategy for incoherent eavesdropping
in Ekert's entanglement based quantum key distribution protocol.
We show that under certain assumptions of symmetry the 
effectiveness of this strategy reduces to that of the original
single qubit protocol of Bennett and Brassard.   
\end{abstract}
\draft
\begin{multicols}{2}

Quantum Key Distribution (QKD) employs quantum features such as
the uncertainty principle and quantum correlations 
to provide for unconditionally secure communications \cite{Mayers}. 
In classical cryptography no mechanism is known for unconditionally 
secure key distribution. The importance of QKD protocols within cryptography
is paramount as they allow for the practical implementation of
unconditionally secure cryptographic algorithms such as the one
time pad \cite{Shannon}.

A quantum cryptographic system can be thought to be constructed of
two main parts; firstly the underlying quantum key distribution
protocol securely establishes a common key between Alice and Bob,
and secondly the message is simply encrypted with the established
key, and transmitted. Theoretical models for quantum key
distribution protocols based on the uncertainty principle have
been analysed by Bennett and Brassard (BB$84$) \cite{Bennett} 
and models based on quantum correlations have 
been proposed by Ekert (E$91$) \cite{Ekert}.

Eavesdropping can be considered as the interception of 
a transmission between Alice and Bob
by an eavesdropper (Eve) and then the application of an 
appropriate measurement to extract information.  
Regardless of the difficulty of interception of the
message, in principle any classical channel can always be
passively monitored, without the legitimate users being aware that
any eavesdropping has taken place.  This is not so for quantum
channels. Any measurement disturbs the transmitted state and hence can be
detected by Alice and Bob. This leads to a trade-off between the amount
of information that Eve can acquire about the transmission and the degree to
which she disturbs it and hence can be detected \cite{Fuchs}.  

In this paper we present an incoherent eavesdropping strategy for
E$91$. The E$91$ protocol can be presented in four main stages: 
\begin{itemize}
\item Entangled photons are distributed to Alice and Bob
who measure them randomly choosing between two fixed measurement
basis. This state is such that in the absence of 
noise or eavesdropping Alice and Bob's 
measurement results are the same if they choose the same basis;
\item Alice and Bob communicate classically their choice of basis
for all pairs and discard all measurement results where different basis 
were used;
\item A subset of the remaining measurement results is publicly 
compared and used to determine the error rate (i.e. where Alice's
and Bob's outcomes do not coincide). This error rate is assumed
to be due to Eve;
\item Finally, if the error rate is below a certain acceptable 
threshold, then the remaining undisclosed outcomes are employed as
the key. Otherwise, the outcomes are discarded and the whole 
process is repeated. 
\end{itemize} 

In our eavesdropping strategy we assume that Eve prepares 
both photons of the entangled pair 
before they are transmitted to Alice and Bob. We do not analyse the
most general attack of this type, but instead focus
on a simpler symmetric attack.

We adopt a conservative view in which Eve prepares a pair of photons 
in an arbitrary state, possibly entangled with other quantum systems. 
These auxiliary quantum systems are known as a probe. 
Without loss of generality~\cite{Uhlmann}, the 
initial joint state of the photon pair 
and the probe can be assumed to be a pure state $\state$ 
of the following form: 
\eq{
\state =\sum_{\alpha,\beta\in\bin} \ketc{\E_{\alpha,\beta}}\ket{\alpha}{+}\ket{\beta}{+},\label{general}
}
where $\{\ket{0}{+},\ket{1}{+}\}$ is a basis for the polarisation 
state of a single photon. The states $\ket{\alpha}{+}$ and $\ket{\beta}{+}$
refer to Alice and Bob's respective photon states. 
The kets for Eve's probe $\ketc{\E_{\alpha,\beta}}$ 
are not necessarily normalised or orthogonal. The only condition on these 
vectors is the overall normalisation of $\state$, leading to:
\eq{
\sum_{\alpha,\beta\in\bin}\scalarc{\E_{\alpha,\beta}}{\E_{\alpha,\beta}}=1.\label{norm1}
}

If we assume that Alice controls the EPR source, 
then the security of the above E$91$ scheme 
is equivalent to the security of BB$84$ protocol \cite{BBM92}. 
This is because Alice first generates an entangled pair locally, keeps
one photon and sends the other to Bob. Therefore only one photon can 
be intercepted and measured by Eve which is thus analogous to BB$84$.

Notice that if Eve is limited only to eavesdropping on the quantum channel 
between the source and Bob, her eavesdropping can be described without 
loss of generality as an unitary interaction between Bob's photon and Eve's probe:
\eqa{
\ketc{F}\ket{0}{+} &\mapsto& \ketc{F_{00}}\ket{0}{+}+\ketc{F_{01}}\ket{1}{+},\\
\ketc{F}\ket{1}{+} &\mapsto& \ketc{F_{10}}\ket{0}{+}+\ketc{F_{11}}\ket{1}{+},
}
where $\ketc{F}$ is the initial state of Eve's probe, and the $\ketc{F_{\alpha,\beta}}$ are again not necessarily normalised or orthogonal, but obey the unitarity conditions:
\eqa{
\scalarc{F_{00}}{F_{00}}+\scalarc{F_{01}}{F_{01}} &=& 1,\label{u1}\\
\scalarc{F_{10}}{F_{10}}+\scalarc{F_{11}}{F_{11}} &=& 1,\label{u2}\\
\scalarc{F_{00}}{F_{10}}+\scalarc{F_{01}}{F_{11}} &=& 0.\label{u3}
}

The resulting global state of the pair and the probe is:
\eq{
|\psi'\rangle = \sum_{\alpha,\beta\in\bin} \frac{\ketc{F_{\alpha,\beta}}}{\sqrt{2}}\ket{\alpha}{+}\ket{\beta}{+}.
}

However, because of the unitarity conditions (Eqn.~(\ref{u1},\ref{u2},\ref{u3})), the global state obtained by Eve when she is only accessing one of the photons is not as general as the state she can obtain when she can access both as in Eqn.~(\ref{general}). 
Therefore this proves that
the eavesdropping strategy in which Eve controls the source (i.e. both photons) 
is potentially stronger 
than the strategy in which she is allowed to eavesdrop on only one photon.

We now turn our attention to a symmetric subclass of incoherent 
attacks against E$91$ in which Eve controls the source. We 
derive the probability that Eve guesses Alice's bit correctly as 
a function of the error probability that Alice and Bob can estimate 
from comparing a fraction of the key bits they obtain.

Let $\{\ket{0}{+},\ket{1}{+}\}$ be a basis for the 
polarisation state of a single photon. We define its 
conjugate basis $\{\ket{0}{\times},\ket{1}{\times}\}$ by $\ket{0}{\times}=(\ket{0}{+}+\ket{1}{+})/\sqrt{2}$ and 
$\ket{1}{\times}=(\ket{0}{+}-\ket{1}{+})/\sqrt{2}$. For 
any $b\in\bases$ we use the shorthand notation $\ket{\alpha \beta}{b}$ 
for the state of a pair of photons in which Alice's photon is in 
state $\ket{\alpha}{b}$ and Bob's photon in the state $\ket{\beta}{b}$. 
We define the Bell basis $\{\ketb{0},\ketb{1},\ketb{2},\ketb{3}\}$ for 
the pair of photons Alice and Bob receive by $\ketb{0}=(\ket{00}{+}+\ket{11}{+})/\sqrt{2}$, $\ketb{1}=(\ket{00}{+}-\ket{11}{+})/\sqrt{2}$, $\ketb{2}=(\ket{01}{+}+\ket{10}{+})/\sqrt{2}$ 
and $\ketb{3}=(\ket{01}{+}-\ket{10}{+})/\sqrt{2}$. In the most 
general incoherent attack, Eve prepares two single photons and a 
probe in any pure state $\state = \sum_{c = 0}^3 \ketc{E_c}\ketb{c}$, where for 
simplicity we use the Bell basis. The kets for the probe, $\ketc{E_c}$ as 
previously mentioned are not necessarily normalised or orthogonal, 
but nevertheless have to obey the overall normalisation relation $\sum_{c=0}^3\scalarc{E_c}{E_c}=1$.  It is straightforward to check that
\eqa{
\state &=& \frac{\ketc{E_0}+\ketc{E_1}}{\sqrt{2}}\ket{00}{+}+\frac{\ketc{E_2}+\ketc{E_3}}{\sqrt{2}}\ket{01}{+}\nonumber\\
&& +\frac{\ketc{E_2}-\ketc{E_3}}{\sqrt{2}}\ket{10}{+}+\frac{\ketc{E_0}-\ketc{E_1}}{\sqrt{2}}\ket{11}{+}\label{inplus}\\
&=& \frac{\ketc{E_0}+\ketc{E_2}}{\sqrt{2}}\ket{00}{\times}+\frac{\ketc{E_1}-\ketc{E_3}}{\sqrt{2}}\ket{01}{\times}\nonumber\\
&& +\frac{\ketc{E_1}+\ketc{E_3}}{\sqrt{2}}\ket{10}{\times}+\frac{\ketc{E_0}-\ketc{E_2}}{\sqrt{2}}\ket{11}{\times}.\label{intimes}
}

Eve sends each of the photons to Alice and Bob. After public communication between 
Alice and Bob, Eve then performs any measurement on her probe. 
We say that an incoherent attack is furthermore \emph{symmetric} if and only if
for any common choice of basis, the probability distribution of Alice's bit is uniform irrespective of
whether or not Alice and Bob share the same bit value.

Given that Eve is restricted to performing an incoherent symmetric attack, the above condition requires that $\Pr(\alpha=0,\beta=0 | \mbox{ basis }b) = \Pr(\alpha=1,\beta=1 |\mbox{ basis } b)$ and $\Pr(\alpha=0,\beta=1 | \mbox{ basis }b) = \Pr(\alpha=1,\beta=0 |\mbox{ basis }b)$ for any choice of basis $b$. This implies that $\Re \scalarc{E_0}{E_1}=\Re\scalarc{E_2}{E_3}=\Re\scalarc{E_0}{E_2}=\Re\scalarc{E_1}{E_3}=0$. 

The probability that Alice and Bob fail to share the same bit value given that they have chosen the same basis $b\in\bases$ reads:
\eqa{
\epsilon_b &=& \Tr\Big[\Id\otimes(\proj{01}{b} + \proj{10}{b})\state\langle\psi| \Big] \\
	&=& \left\{ \begin{array}{cl} z + t & \mbox{ if } b=+, \\ y+t & \mbox{ if } b=\times\end{array}\right.
}
where the identity operator acts on the Hilbert space of the probe and where we have used the shorthand notation $x=\scalarc{E_0}{E_0}$, $y=\scalarc{E_1}{E_1}$, $z=\scalarc{E_2}{E_2}$ and $t=\scalarc{E_3}{E_3}$.

We assume that, after public discussion between Alice and Bob, Eve is aware of 
their chosen basis, and whether or not their measurements yielded 
the same bit value. From this data Eve's objective is to guess Alice's bit value 
$\alpha$. Assuming that Alice and Bob chose the basis $b=+$ and that their 
measurement returned the {\em same} bit value, we see from Eqn.~(\ref{inplus}) 
that Eve has to find out whether her probe is in the state $\frac{\ketc{E_0}+\ketc{E_1}}{\sqrt{x+y}}$, corresponding to $\alpha=0$ or 
in the state $\frac{\ketc{E_0}-\ketc{E_1}}{\sqrt{x+y}}$ corresponding to $\alpha=1$.  
Given a quantum system that is equally likely to be in either the state $\ketc{\eta}$ or 
$\ketc{\chi}$, it is known that the optimal probability of guessing 
this state correctly is $P_c = \frac{1}{2}+\frac{1}{2}\sqrt{1-|\scalarc{\eta}{\chi}|^2}$
~\cite{Hel76}. In our case, 
Eve's probability of guessing correctly is:
\eq{
P_{c| \share, b=+} = \frac{1}{2}+\frac{1}{2}\sqrt{1-\left(\frac{x-y}{x+y}\right)^2}.
}

If Alice and Bob chose the basis $b=+$, but that their 
measurements returned {\em different} bit values, Eve has to find out 
whether her probe is in the state $\frac{\ketc{E_2}+\ketc{E_3}}{\sqrt{y+z}}$ or $\frac{\ketc{E_2}-\ketc{E_3}}{\sqrt{y+z}}$. Eve's probability of guessing correctly is now:
\eq{
P_{c| \nshare, b=+} = \frac{1}{2}+\frac{1}{2}\sqrt{1-\left(\frac{z-t}{z+t}\right)^2}.
}

Similar calculations in the conjugate basis give:
\eqa{
P_{c| \share, b=\times} &=&  \frac{1}{2}+\frac{1}{2}\sqrt{1-\left(\frac{x-z}{x+z}\right)^2}\,\mbox{ and,} \\
P_{c| \nshare, b=\times} &=& \frac{1}{2}+\frac{1}{2}\sqrt{1-\left(\frac{y-t}{y+t}\right)^2}.
}
 
Given that the choice of the basis $b$ is uniformly distributed, the marginal 
probability that Eve guesses Alice's bit correctly is:
\eqa{
P_c &=& \frac{1-\epsilon_+}{2}P_{c| \share, b=+} + \frac{\epsilon_+}{2}P_{c| \nshare, b=+}\nonumber\\
&& + \frac{1-\epsilon_\times}{2}P_{c| \share, b=\times} + \frac{\epsilon_\times}{2}P_{c| \nshare, b=\times}\\
&=& \frac{1}{2}+\frac{1}{2}\left(\sqrt{xy}+\sqrt{zt}+\sqrt{xz}+\sqrt{yt}\right)\\
&=& \frac{1}{2}+\frac{1}{2}\left(\sqrt{1-\epsilon_+-x}+\sqrt{1-\epsilon_\times-x}\right)\nonumber\\
&& \times\left(\sqrt{x}+\sqrt{x+\epsilon_++\epsilon_\times-1}\right), 
}
where we have used the normalisation condition $x+y+z+t=1$. Since $y$, $z$ and $t$ are non negative numbers, the domain for $x$ is $[1-\epsilon_+-\epsilon_\times, 1-\max(\epsilon_+,\epsilon_\times)]$. 

Now given the average error probability $\overline{\epsilon}=\frac{\epsilon_++\epsilon_\times}{2}$, the above probability of a correct guess is maximal when $\epsilon_+ = \epsilon_\times=\overline{\epsilon}$. In other words, given the average error probability, the eavesdropping strategy that maximises the average probability of correct guess is the one for which the error probability is independent of the choice of basis. The resulting probability of correct guess:
\eq{
P_c = \frac{1}{2}+\sqrt{1-\overline{\epsilon}-x}\left(\sqrt{x}+\sqrt{2\overline{\epsilon}+x-1}\right)
}
reaches its maximum for $x=1-2\overline{\epsilon}+\overline{\epsilon}^2$ at which point:
\eq{
P_c = \frac{1}{2} + \sqrt{\overline{\epsilon}(1-\overline{\epsilon})}.
}

The above formula is identical to the formula obtained by Cirac and Gisin~\cite{CG97} in their study of the security of BB$84$ against a class of incoherent attacks with similar conditions for symmetry (see Fig. 1). The fact that our incoherent symmetric eavesdropping strategy 
has effectively the same power as the BB$84$ incoherent symmetric 
eavesdropping strategy is surprising. This is particularly so given that the maximally entangled state $|00\rangle_++|11\rangle_+$ has the following interesting property:
\eq{
A\otimes I (|00\rangle_++|11\rangle_+)= I\otimes A^{T}(|00\rangle_++|11\rangle_+), 
}
where $A$ is any operator. This implies that any operation on one qubit
can be executed remotely by performing the transpose of that operation
on the other qubit, providing that the state of the two qubits is 
maximally entangled. In spite of this, there is no operator $C$
such that 
\eq{
A\otimes B (|00\rangle_++|11\rangle_+)= I\otimes C (|00\rangle_++|11\rangle_+), 
}
which is why the two qubit attack is strictly more general than the single
qubit attack. This is of course because the operation represented by $B$ 
will in general result in a non-maximally entangled state which no longer 
has the property of remote execution of operations. 
 
Note that the above result can also be expressed as a trade-off between the error
rate and the mutual information obtained by Eve about the nature of the 
state communicated by Alice to Bob. In this case we have a situation 
equivalent to a binary symmetric channel so that the mutual information
is the capacity of that channel and is given by \cite{Shannon}
\eq{
I = 1 + P_c \log P_c + (1-P_c) \log (1-P_c) .
}  
As before, the greater the mutual information gained by Eve the greater 
the (detectable) error rate induced in the communication.

In this paper we have analysed a symmetric incoherent eavesdropping 
strategy in E$91$ quantum key distribution protocol. We have assumed
that Eve controls preparation of the entangled photons. In spite
of this, we have found that there is no benefit for the purpose
of eavesdropping when compared to symmetric incoherent
eavesdropping in BB$84$. We hope that this stimulates further
investigation into entanglement based quantum cryptography.

This work was supported in part by the European TMR Research
Network ERP-4061PL95-1412, Hewlett Packard and Elsag plc. We gratefully acknowledge intersting discussion with Hans Briegel, Artur Ekert, Norbert L\"utkenhaus and Dominic Mayers.  

\begin{figure}[ht]
\begin{center}
\hspace{0mm}
\epsfxsize=7.5cm
\epsfbox{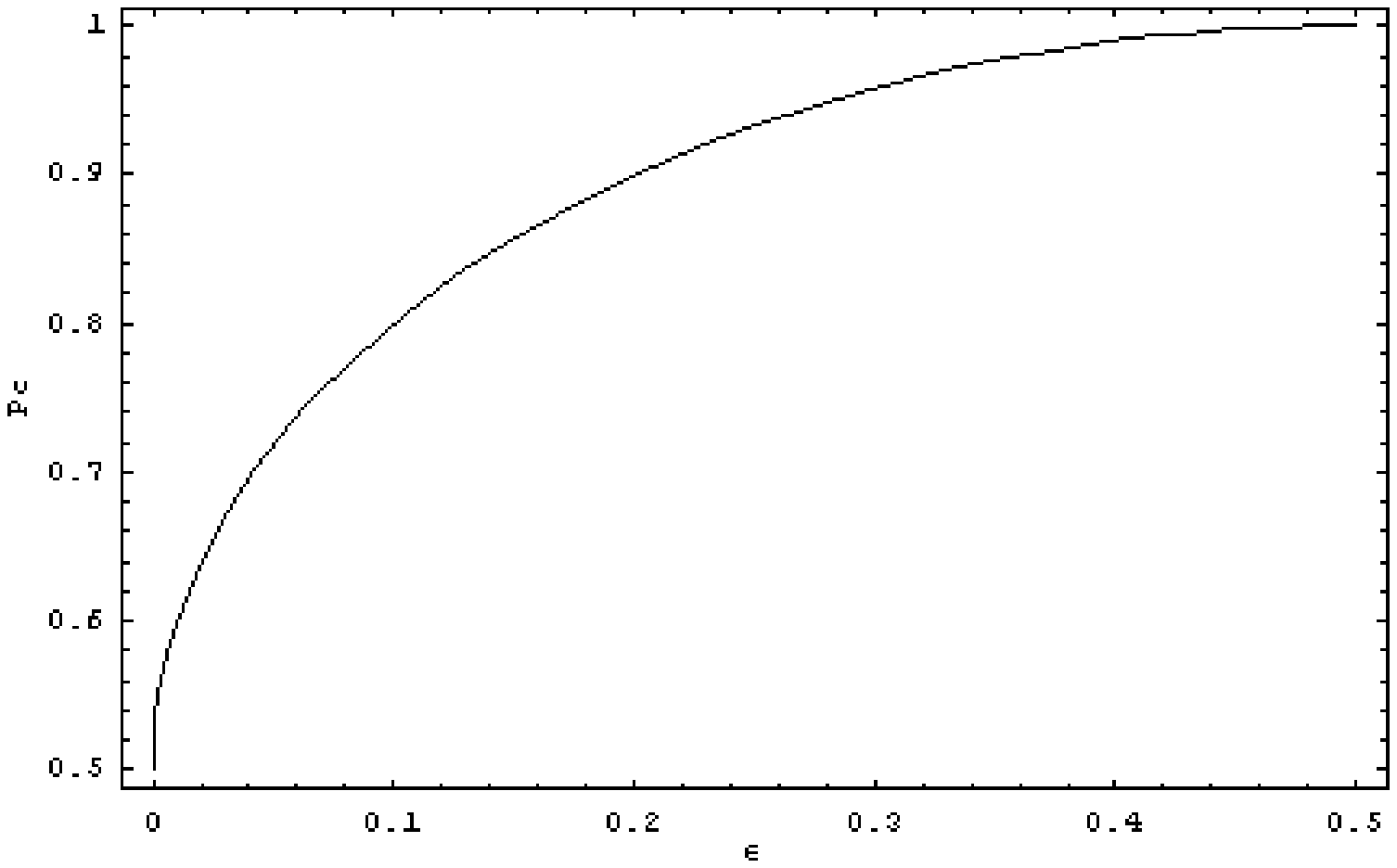}\\[0.2cm]
\begin{caption}
{\narrowtext The figure shows the efficiency of the symmetric 
two qubit attack. The probability of correct guess is plotted against the
error rate induced by eavesdropping.}
\end{caption}
\end{center}
\end{figure}

\end{multicols}

\end{document}